\documentclass{ctr}

\usepackage{ctrfont}
\usepackage{natbib}

\usepackage{url}
\usepackage{amsmath}
\usepackage{amssymb}

\usepackage[dvips]{graphicx}
\usepackage{psfrag}
\usepackage{epsfig}
\usepackage{subfigure}

\newcommand{\ubar}{\overline{u}}
\newcommand{\rst}{\overline{u'_iu'_j}}
\newcommand{\dif}{\mathrm{d}}
\newcommand{\Dif}{\mathrm{D}}

\title{A return to eddy viscosity model for epistemic UQ in RANS closures}
\shorttitle{A return to eddy viscosity}
\author{W.N. Edeling,  G. Iaccarino \and P. Cinnella\footnote{Arts et 
M\'{e}tiers ParisTech, Paris, France.}}
\shortauthor{Edeling et al.}

\begin{document}


\maketitle

\section{Motivation and objectives} 


Due to their computational tractability, closure models based on the 
Reynolds-Averaged Navier-Stokes (RANS) equations remain a widely used option 
for the computation of turbulent flow fields. However, the numerous assumptions 
made in their derivation and calibration result in models subject to unknown 
degrees of parametric and epistemic model-form uncertainty. This in turn leads 
to flow-dependent performance and ultimately to predictions which can be 
trustworthy for certain regions of a flow domain, yet highly erroneous in
others. 

Various attempts have been made to quantify the uncertainty in RANS closures. 
Bayesian inference can be used to obtain stochastic estimates of the closure 
coefficients, (see, e.g., \citep{edeling2014bayesian}). A limitation is that 
just varying the coefficients does not challenge the underlying Boussinesq 
hypothesis. Also, the obtained posterior distributions on the coefficients have 
no direct spatial dependence. They are applied uniformly throughout the 
flow domain, regardless of the degree of local model failure.

Recently, Machine-Learning (ML) algorithms have also found application to the 
problem. For instance \citep{zhang2015machine} have applied neural nets and 
Gaussian processes to directly infer new so-called adjustment terms from 
high-fidelity data. These terms are added to an existing closure model 
with the intention of correcting for the RANS bias. As they are trained on local 
input features, the adjustment term is able to vary spatially. 

Although promising results have been obtained, (see, e.g., 
\citep{wu2016physics}), ML techniques are not without their drawbacks. 
Depending on the chosen ML algorithm, the inverse problem can be 
high-dimensional. Also, taking a neural network as an example, the manner in 
which it arrives at the complex input-output relation is not easy to 
interpret in physical terms and therefore does not necessary lead to increased 
confidence. Furthermore, the process is entirely dependent on the training and 
calibration data used and therefore possibly is of limited universality. 
And although it can approximate complex functions, it yields no analytical form to 
study the structure of the model discrepancy.

We will build upon the work of \citep{emory2013modeling}, in which 
perturbations in the eigenvalues of the anisotropy tensor are introduced in 
order to bound a Quantity-of-Interest (QoI) based on limiting states of 
turbulence. To make the perturbations representative of local flow features, 
we introduce two additional transport equations for linear combinations of these 
aforementioned eigenvalues. The model form is inspired by the LAG model of 
\citep{olsen2001lag}, which perturbs the eddy-viscosity at locations where a 
failure of the Boussinesq hypothesis might reasonably be anticipated. The 
location, magnitude and direction of the eigenvalue perturbations are now 
governed by the model transport equations. The general behavior of our 
discrepancy model is determined by two coefficients, resulting 
in a low-dimensional Uncertainty Quantification (UQ) problem. We will 
furthermore show that the behavior of the model is intuitive and rooted in the 
physical interpretation of misalignment between the mean strain and Reynolds 
stresses. 

In this paper we will focus on estimating prediction intervals on QoIs. 
Obtaining data-driven distributions within those intervals in a Bayesian setting 
is the subject of ongoing research.  In this sense, the intervals of this paper 
might be considered as the results of propagating uniform prior distributions on 
the two coefficients. 

The structure of the paper is as follows. In Section 2 we discuss the 
representation of the Reynolds stress anisotropy, followed by a 
section on the model transport equations and their properties. Section 4 covers 
the obtained intervals on the QoI and finally, we close by 
presenting our conclusions in Section 5.



\section{Reynolds stress anisotropy and the barycentric map}

A RANS closure employed by many turbulence models is the Boussinesq 
hypothesis, which reads
\begin{align}
 \rst  \approx \frac{2}{3}k\delta_{ij} - 2\nu_TS_{ij}.
 \label{eq:boussinesq}
\end{align}
Here, $k$ is the turbulent kinetic energy, $\delta_{ij}$ the Kronecker delta, 
and $S_{ij}$ is the mean strain rate tensor, defined as 
$S_{ij}:=\frac{1}{2}\left({\partial \ubar_i}/{\partial x_j} + 
{\partial \ubar_j}/{\partial x_i}\right)$ in the case of incompressible 
flow. The scalar quantity $\nu_T$ is the eddy-viscosity, which must be computed 
by means of a chosen turbulence model, e.g. the $k-\varepsilon$, $k-\omega$ or 
Spalart-Allmaras model \citep{wilcox1998turbulence}. Although we will use the
$k-\varepsilon$ model, all closure models have their own specific mathematical
form, closure coefficients and corresponding uncertainties, ultimately leading
to an uncertain $\nu_T$ \citep{edeling2014bayesian}. Instead of focusing on 
$\nu_T$, we will inject uncertainty directly into $\rst$ by adding a tensorial 
discrepancy model $E_{ij}$ to the right-hand side of Eq. (\ref{eq:boussinesq}), 
leaving any underlying baseline turbulence model unperturbed. To this end, let 
us define the normalized anisotropy tensor $b_{ij}$ as
\begin{align}
 b_{ij}:=\frac{\rst}{2k} - \frac{1}{3}\delta_{ij};\quad
 -\frac{1}{3}\leq b_{\alpha\alpha} \leq \frac{2}{3};\quad
 -\frac{1}{2}\leq b_{\alpha\beta} \leq \frac{1}{2};
 \label{eq:b_ij}
\end{align}
\noindent
which is a symmetric, deviatoric (zero trace) tensor. Greek subscripts are 
excluded from the summation convention. Tensor invariants of $b_{ij}$ are a 
useful tool to study the anisotropy. These invariants are defined as
\begin{align}
 \mathrm{I}_b = b_{ii} \quad 
 \mathrm{II}_b = b^2_{ii} = b_{ij}b_{ji}\quad 
 \mathrm{III}_b = b^3_{ii} = b_{ij}b_{jk}b_{ki}.
 \label{eq:invar1}
\end{align}
\noindent
Since $b_{ij}$ is deviatoric ($\mathrm{I}_b = 0$), only two invariants are
needed to characterize the state of anisotropy. The invariants in Eq. 
(\ref{eq:invar1}) 
are non-linear functions of the eigenvalues of $b_{ij}$, (see, e.g., 
\citep{lumley1977return}). More recently, other invariant 
measures have been proposed by \citep{banerjee2007presentation} which 
have a linear relationship with the eigenvalues of $b_{ij}$. As our framework 
will involve inverting the relation between the invariants and the eigenvalues, 
these linear measures are preferred. 

Banerjee expresses the anisotropy tensor as a convex combination of three 
basis tensors $\hat{b}_{1c}$, $\hat{b}_{2c}$, $\hat{b}_{3c}$, i.e.,
\begin{align}
 \hat{b}_{ij} = C_{1c}\hat{b}_{1c} + C_{2c}\hat{b}_{2c} + C_{3c}\hat{b}_{3c}.
 \label{eq:b_ij_decomp}
\end{align}
\noindent
These basis tensors represent the three limiting states of componentality 
(relative strengths of components in $\rst$), i.e., they represent one-, two- 
and three-component turbulence. The modified notation $\hat{b}_{ij}$ 
represents the anisotropy tensor in principal axes,
\begin{align}
 \hat{b}_{ij} = \mathrm{diag}\left(\lambda_1, \lambda_2, \lambda_3\right),
 \label{eq:b_ij_pcs}
\end{align}
\noindent
where $\lambda_1 \geq \lambda_2 \geq \lambda_3$.
%
The basis tensors are defined as
\begin{align}
 \hat{b}_{1c} = \mathrm{diag}\left(2/3, -1/3, -1/3\right),\;\;
 \hat{b}_{2c} = \mathrm{diag}\left(1/6, 1/6, -1/3\right),\;\;
 \hat{b}_{3c} = \mathrm{diag}\left(0, 0, 0\right),
\end{align}
\noindent
where $\hat{b}_{3c}$ corresponds to the isotropic limit.
%
%
%
%
%
%

From Eq. (\ref{eq:b_ij_decomp}) it is clear that the coefficients $C_{1c}$, 
$C_{2c}$ and $C_{3c}$ measure how close $\hat{b}_{ij}$ is to any of the three 
limiting states. Since Eq. (\ref{eq:b_ij_decomp}) is required to be a convex 
combination, we have by definition:
\begin{align}
 C_{1c} + C_{2c} + C_{3c} = 1 \label{eq:C_sum_to_1}\;\;\mathrm{and}\;\;
 C_{1c} \geq 0\quad C_{2c} \geq 0\quad C_{3c} \geq 0.
\end{align}
\noindent
Requiring further that
\begin{align}
\mathrm{At\;\;1C:}\quad C_{1c} = 1, \quad C_{2c} = C_{3c} = 0 \nonumber\\
\mathrm{At\;\;2C:}  \quad C_{2c} = 1, \quad C_{1c} = C_{3c} = 0 \nonumber\\
\mathrm{At\;\;3C:}  \quad C_{3c} = 1, \quad C_{1c} = C_{2c} = 0
\end{align}
\noindent
yields the following linear (inverse) relationship between coefficients and 
the eigenvalues of $\hat{b}_{ij}$
\begin{align}
\begin{tabular}{ccc}
 $C_{1c} = \lambda_1 - \lambda_2$ & $\Rightarrow$ &
 $\lambda_1 = C_{1c} + \frac{C_{2c}}{2} + \frac{C_{3c}}{3} - \frac{1}{3}$
 \\
 $C_{2c} = 2\left(\lambda_2 - \lambda_3\right)$ & $\Rightarrow$ &
 $\lambda_2 = \frac{C_{2c}}{2} + \frac{C_{3c}}{3} - \frac{1}{3}$ 
 \\
 $C_{3c} = 3\lambda_3 + 1$  & $\Rightarrow$ &
 $\lambda_3 = \frac{C_{3c}}{3} - \frac{1}{3}$
\end{tabular}.
 \label{eq:bary_coefs}
\end{align}
\noindent
Due to Eq. (\ref{eq:C_sum_to_1}), again only two coefficients are needed to 
quantify the state of anisotropy. 

\begin{figure}
\centering
\includegraphics[scale=0.5]{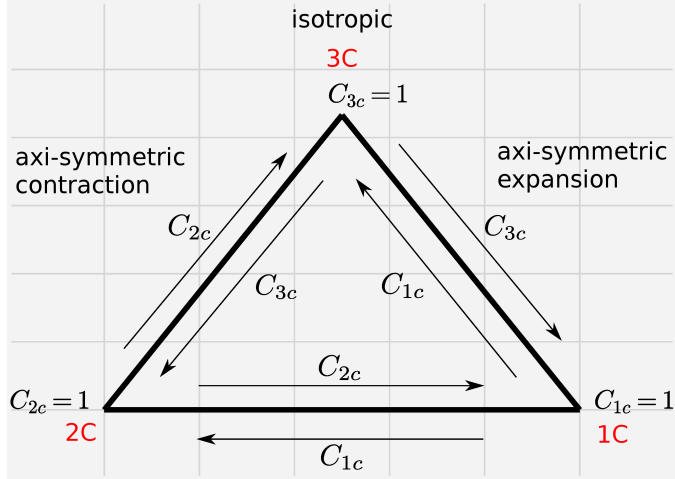}
\caption{The barycentric map with the one-, two- and three-component corners, 
and the variation of its coefficients. Arrows indicates directions along which 
the corresponding coefficient linearly decreases from one towards
zero.\label{fig:bary}}
\end{figure}

To visualize the nature of the anisotropy implied by the invariants, a 
barycentric map can be defined as \citep{banerjee2007presentation}
\begin{align}
 x_b = C_{1c}x_{1c} + C_{2c}x_{2c} + C_{3c}x_{3c}, \nonumber\\
 y_b = C_{1c}y_{1c} + C_{2c}y_{2c} + C_{3c}y_{3c}.
 \label{eq:bary_map}
 \end{align}
 \noindent
Here, $(x_{1c}, y_{1c})$, $(x_{2c}, y_{2c})$ and $(x_{3c}, y_{3c})$ are the 
three corner points corresponding to the limiting states of componentality. 
They can be chosen arbitrarily, but are commonly set to the corner points of an 
equilateral triangle. In Figure \ref{fig:bary} the barycentric map is depicted 
along with the variation of the coefficients along its edges. Each point ${\bf 
x}\in\mathbb{R}^N$ in the spatial flow domain has its own coefficients and thus 
can be mapped to a location in the barycentric map using 
Eq. (\ref{eq:bary_map}). 
All possible realizable states of $\rst$ are contained within the borders of 
the barycentric map (see Section \ref{sec:rea} for a more detailed discussion).


\section{Reynolds stress perturbation}

\subsection{Anisotropy tensor decomposition}

The eigen decomposition of the anisotropy tensor reads
\begin{align}
 b_{ij} = v_{ik}\Lambda_{kl}v_{jl},
\end{align}
\noindent
where $v_{ij}$ are the eigenvectors and 
$\Lambda_{ij}:=\mathrm{diag}\left(\lambda_1, \lambda_2, \lambda_3\right)$ is the 
diagonal matrix of eigenvalues. Hence, the 
Reynolds stress tensor can be written as
\begin{align}
 \rst = 2k\left(\frac{1}{3}\delta_{ij} + v_{ik}\Lambda_{kl}v_{jl}\right)
 \label{eq:rst_decomp}
\end{align}
\noindent
Here, the turbulent kinetic energy, eigenvectors and eigenvalues represent the 
magnitude, orientation and shape of the Reynolds stress respectively. 
In \citep{mishra2016sensitivity}, the authors have attempted to quantify the 
uncertainty in RANS closures due to coarse graining, to guide the perturbations 
to the turbulent kinetic energy. For an approach perturbing the eigenvectors 
$v_{ij}$, see \citep{mishra2016enveloping}. 

\cite{emory2011modeling} used 
Eq. (\ref{eq:rst_decomp}) to perturb the baseline Reynolds stress tensor 
by replacing the eigenvalue matrix $\Lambda_{ij}$ with a perturbed matrix 
$\Lambda^*_{ij}$. Starting from the baseline result in the barycentric map, 
perturbations were made towards the three corners of the barycentric map. At 
these new locations the perturbed tensor $\Lambda^*_{ij}$ could be calculated 
through Eq. (\ref{eq:bary_coefs}). This resulted in three additional code 
evaluations, the 
result of which were used to bound the QoI based on the limiting states of 
componentality. To determine where in the spatial domain the perturbations 
should be applied, specialized sensors based on wall-distance 
\citep{emory2011modeling} or streamline curvature 
\citep{gorle2012epistemic}
were developed to flag regions where the turbulence model assumptions were most 
likely to be invalid. Within those regions flagged by the sensors, the 
perturbations in the direction of the corners were either homogeneous and 
user-specified or determined by a separate model, e.g., an error model based on 
log-normal distributions of observed discrepancies between RANS and DNS. We 
refer to \citep{emory2013modeling} for more details. 


\subsection{Invariant transport model}

The aim of the current article is to perturb the eigenvalues of $b_{ij}$ in a 
more general fashion, representative of spatial features without the use of 
separate sensors. To this end, we define two additional model transport 
equations for the invariant coefficients $C_{1c}$ and $C_{2c}$. Transport 
equations for tensor invariants of $b_{ij}$ already exist in the context of 
return-to-isotropy models. For instance, in the case of decaying homogeneous 
anisotropic turbulence, Rotta's model \citep{rotta1951statistische} can be 
used
as a starting point to arrive at the following equation for the evolution of 
the tensor invariants of $b_{ij}$ of Eq. (\ref{eq:invar1})
\begin{align}
 \frac{\dif b^n_{ii}}{\dif t} = -n\left(C_R - 
1\right)\frac{\varepsilon}{k}b^n_{ii},
\label{eq:rotta}
\end{align}
\noindent
where $C_R = 1.8$ \citep{pope2001turbulent}. 

However, we deal with a baseline eddy viscosity model which can be erroneous 
at certain locations, but which is able to yield accurate predictions outside 
these regions. Therefore, the transport equations for $C_{1c}$ and 
$C_{2c}$ should describe a return-to-eddy-viscosity when the Boussinesq 
hypothesis is expected to be valid.

A class of turbulence models which does so includes the LAG-type models  
\citep{olsen2001lag}, \citep{lillard2012lagrst}, which add extra transport 
equations 
to account for non-equilibrium effects. Generally speaking, a flow is in
equilibrium when the turbulent time scale is much smaller than
the mean flow time scale, allowing the turbulence to react quickly to changes in
the mean flow. In these types of circumstance, a direct proportionality between
the Reynolds stress tensor and the mean strain rate is a reasonable assumption.
However, when this is not the case, there is a lag in the response of
the turbulence to changes in the mean flow. This lag cannot be accounted for by
eddy-viscosity models since $\rst$ reacts immediately to changes in $S_{ij}$. To
rectify this behavior, \cite{lillard2012lagrst} add the
following six transport equations
\begin{align}
 \frac{\Dif\rst}{\Dif t} = 
a_0\frac{\epsilon}{k}\left(\rst^{\left(bl\right)} - \rst\right),
 \label{eq:lag}
\end{align}
\noindent
where $\rst^{(bl)}$ is the baseline Reynolds stress tensor computed 
with the Boussinesq hypothesis. The central idea of Eq. (\ref{eq:lag}) is to
modulate the six components of $\rst$, such that the
turbulence model no longer responds too rapidly to changes in the mean flow.
Along a steamline, a $\rst$ component goes to its baseline value with a time
scale $\epsilon/k$ computed from the underlying baseline model. 

The rate of the return to eddy viscosity can thus be controlled by the value of
$a_0$, for which the proposed value is 0.35 \citep{olsen2001lag}. As is the
case with most turbulence models, this value is ad hoc and not supported by
theory. In \citep{lillard2012lagrst}, the model response is evaluated at two
different values of $a_0$, but validated only on flat-plate experimental
velocity data. Moreover, it was recently argued in
\citep{spalart2015philosophies} that equilibrium is not a well defined
concept, and that near the wall the deep anisotropy is the cause of RANS model
failure, rather than a lag in the response to changes in $S_{ij}$. In any case,
here we use the ideas of Eqs. (\ref{eq:rotta})-(\ref{eq:lag}) to perturb the
solution away from the Boussinesq hypothesis,
possibly towards a more anisotropic state, in a low-dimensional fashion. To that
end, we propose the following model transport equations for $C_{1c}$ and
$C_{2c}$:
\begin{align}
\frac{\mathrm{D}C_{1c}}{\mathrm{D} t} = 
a_{1c}\frac{\epsilon}{k}\left(C_{1c}^{(bl)} - 
C_{1c}\right) + 
\frac{\partial}{\partial x_i}\left((\nu + 
\frac{\nu_T}{\sigma_{1c}})\frac{\partial C_{1c}}{\partial x_i}\right) 
\nonumber\\
\frac{\mathrm{D} C_{2c}}{\mathrm{D}t} = 
a_{2c}\frac{\epsilon}{k}\left(C_{2c}^{(bl)} - C_{2c}\right) + 
\frac{\partial}{\partial x_i}\left((\nu + 
\frac{\nu_T}{\sigma_{2c}})\frac{\partial C_{2c}}{\partial x_i}\right).
\label{eq:model}
\end{align}
\noindent
The  $C^{(bl)}_{1c}$, $C^{(bl)}_{2c}$ are the coefficients computed using the 
eigenvalues of the Boussinesq anisotropy tensor, i.e., from
\begin{align}
 b^{(bl)}_{ij} = -\frac{\nu_T}{k}S_{ij}.
 \label{eq:b_ij_bous}
\end{align}
\noindent
We will not commit to a single value for the coefficients of the LAG term (as
optimal values are likely to be flow dependent
\citep{edeling2014bayesian}), and instead consider a range of coefficients.
Furthermore, we will examine the effect of lagging one equation more than the
other in order to cover a wider area in the barycentric map, thereby
incorporating more of the uncertainty in the eigenvalues of $b_{ij}$. 

We define the boundary conditions following an approach similar to the 
original LAG model \citep{olsen2001lag}. At a solid smooth wall, we set the 
values of $C_{1c}$ and $C_{1c}$ equal to the baseline values $C^{(bl)}_{1c}$ 
and $C^{(bl)}_{2c}$. If the turbulence model equations are integrated down to 
the wall using damping functions, this essentially sets the values equal to 
zero. However, if wall functions are used, non-zero values of $C_{1c}$ and 
$C_{2c}$ are obtained. At an inflow boundary, it is assumed that the freestream 
turbulence is isotropic, i.e., $\rst = {2}/{3}k_{\infty}\delta_{ij}$, 
yielding $C_{1c} = C_{2c} = 0$. Outflow boundary values are extrapolated 
from internal cells.

The decision of where, 
how much and in what direction to perturb the eigenvalues of $b_{ij}$ is now 
determined by Eq. (\ref{eq:model}), and does not require any user input besides 
the 
specification of the coefficients $a_{1c}$, $a_{2c}$, $\sigma_{1c}$ and 
$\sigma_{2c}$. In cases where the left-hand side of Eq. (\ref{eq:model}) is 
zero 
(e.g., a quasi one-dimensional boundary layer), the diffusion term is needed 
to obtain a result different from the baseline state. Otherwise, one might 
choose 
to neglect this term, in line with the orginal LAG model. For now we 
include diffusion, but our focus is on the effect of the LAG term. As will be 
discussed in Section \ref{sec:lag_coefs}, writing transport 
equations for the barycentric coefficients rather than the Reynolds stresses
yields a model that behaves intuitively with the choice of LAG coefficients
$a_{1c}$ and $a_{2c}$. Also, since we keep the closure coefficients of the
underlying baseline model fixed, these LAG coefficients are the only unknowns.
Determining the effect of these coefficients on the QoI constitutes a 
low-dimensional UQ problem for which many off-the-shelf
techniques exist, e.g. stochastic collocation or polynomial chaos methods
\citep{eldred2009comparison}.


\subsection{Effect of LAG coefficients \label{sec:lag_coefs}}

\begin{figure}
 \centering
 \includegraphics[scale=0.65]{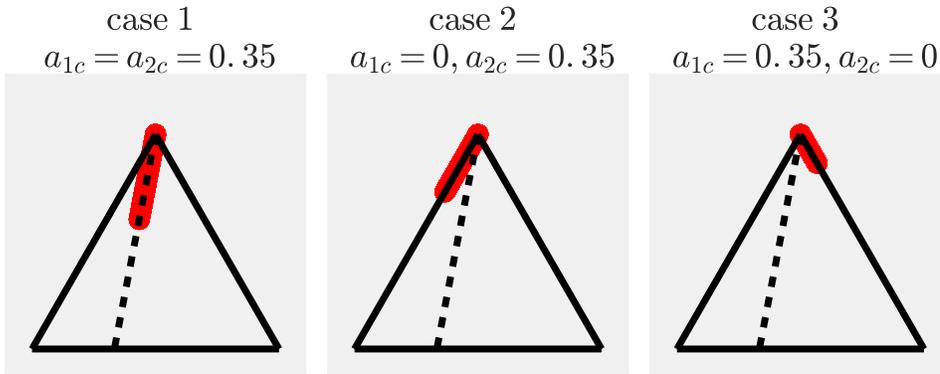}
 \caption{The trajectories in the barycentric map of the three coefficient 
cases. The dashed line denotes the plane strain line where $\lambda_2 = 0$. 
\label{fig:bary_case123}}
\end{figure}

We specify the default values of the coefficients in (\ref{eq:model}) as
\begin{align}
 a_{1c} = a_{2c} = 0.35  \quad\quad \sigma_{1c} = \sigma_{2c} = 1.0.
\end{align}
\noindent
As previously explained, these are ad hoc choices and we do not claim 
universality on these values. The lag coefficients $a_{1c}$ and 
$a_{2c}$ determine the rate 
of return to eddy viscosity. Higher values result in $C_{1c}$ or $C_{2c}$ 
distributions  that more closely follow the Boussinesq result. Let us define 
three cases that represent limiting trajectories:
\begin{enumerate}
 \item[ Case 1 (plane-strain):] $a_{1c} = a_{2c} = 0.35$,
 \item[ Case 2 (axi-symmetric contraction):] $a_{1c} = 0$, $a_{2c} = 0.35$,
 \item[ Case 3 (axi-symmetric expansion):] $a_{1c} = 0.35$, $a_{2c} = 0$.
 \label{eq:coefs}
\end{enumerate}
\noindent
The resulting trajectories in the barycentric map, computed on a flat plate 
case, are shown in Figure \ref{fig:bary_case123}. 
Before discussing these results, let us first note that the  
unperturbed baseline result will lie on top of the plane 
strain line for this flow case. A model will follow the plane strain turbulence 
line when at least one eigenvalue of $b_{ij}$ is zero 
\citep{banerjee2007presentation}. 

When $a_{1c}=a_{2c}$ (case 1), both transport equations in Eq. (\ref{eq:model}) 
return to the Boussinesq solution at the same rate. As a result, the trajectory 
in the barycentric map still lies along the plane strain line (see Figure 
\ref{fig:bary_case123}). Along this line the distribution is 
different than that of the unperturbed model, depending on the values of 
$a_{1c}$ and $a_{2c}$. 

To observe larger deviations from the results of the 
baseline model, unequal LAG coefficients are required. In the second limiting 
case we enforce a homogeneous solution $C_{1c}=  0$ by setting $a_{1c} = 0$.
This forces the trajectory to follow the axi-symmetric contraction border (see 
again Figure \ref{fig:bary_case123}). Similarly, setting $a_{2c} = 0$ results 
in a trajectory along the axi-symmetric expansion border, where $C_{2c} = 0$. 

%
%

\subsection{Model behavior at 3C}

When $a_{1c} = a_{2c} = 0.0$, all points in the flow domain will be confined to 
the 3C corner of the barycentric map. The behavior of the model in this limit 
will be determined by the representation of the production term in the $k$ 
equation. The exact production term reads
\begin{align}
 \mathcal{P} := -\rho\rst\frac{\partial\bar{u}_i}{\partial x_j},
 \label{eq:P_exact}
\end{align}
\noindent
which is modeled as 
$\mathcal{P}:=\rho\nu_T\left({\partial\bar{u}_i}/{\partial x_j}\right)^2$ 
in the $k-\varepsilon$ model \citep{wilcox1998turbulence}. If the perturbed 
Reynolds stress tensor is used in Eq. (\ref{eq:P_exact}), the flow becomes 
laminar when all points are perturbed to the 3C corner (since the off-diagonal 
component is the dominant contribution to $\mathcal{P}$ in this case). If the 
modeled 
production term is retained, the flow becomes 
isotropic with non-zero $k$. Employing the latter option has a number of 
advantages. First, from a UQ perspective one might be interested in the 
uncertainty of a given baseline model, and will therefore not alter the 
mathematical form of this model before performing the UQ analysis. Second, 
laminarizing the flow can lead to convergence issues. For investigating
the behavior of model Eq. (\ref{eq:model}), we wish to retain the ability of 
perturbing all the way to the 3C corner, and hence we keep the Boussinesq 
production term. That said, it should be noted that when one relaxes this 
requirement, the use of Eq. (\ref{eq:P_exact}) can improve the $k$ 
predictions \citep{gorle2012epistemic}.


\subsection{Spatial distribution of perturbations}

\begin{figure}
\centering
\includegraphics[scale=0.5]{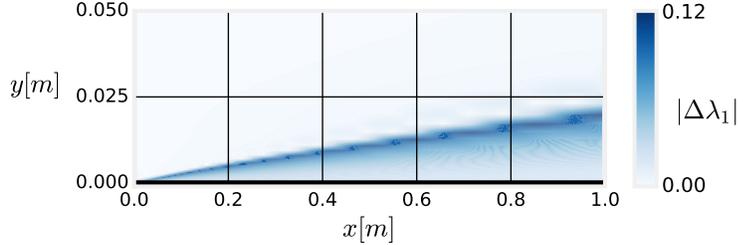}
\caption{A contour plot of $|\Delta\lambda_1|:= |\lambda_1 - \lambda_1^{(eq)}|$
for $a_{1c} = a_{2c} = 0.35$.\label{fig:locality}}
\end{figure}

The extend of baseline model failure will depend upon the 
local flow physics, (see e.g., \citep{ling2015evaluation}) for a detailed 
discussion of localized RANS failure. Ideally, the location, magnitude and 
direction of the perturbations should reflect this non-uniform nature of the 
RANS error. As mentioned, one may employ specialized marker functions based on
local flow features to restrict the perturbations to regions where failure of
the baseline model can reasonably be expected \citep{emory2013modeling}, 
\citep{gorle2012epistemic}. Besides determining the magnitude and 
direction, the
LAG term also doubles as a marker function which flags regions where the
Boussinesq assumption is likely to be invalid \citep{lillard2012lagrst}. As an
example, consider the spatial distribution of $|\lambda_1 - \lambda_1^{(eq)}|$ 
for a flat plate flow shown in Figure \ref{fig:locality}. Here, 
$\lambda_1^{(eq)}$ is the largest eigenvalue of Eq. (\ref
{eq:b_ij_bous}). Only close to the wall do we find any significant departure 
from the Boussinesq value of the largest eigenvalue. Away from the wall, where 
the flow is essentially uniform, we find no perturbations.

\subsection{Realizability \label{sec:rea}}

To obtain physically feasible values for the components of the Reynolds stress 
tensor, the realizability constraints first defined by
\citep{schumann1977realizability} must hold, i.e.
\begin{align}
 \overline{u'_{\mu}u'_{\mu}} \geq 0\quad\quad
 (\overline{u'_\mu u'_\nu})^2 \leq 
\overline{u'_\mu u'_\mu}\;\overline{u'_\nu u'_\nu}\quad\quad
 \det\left(\rst\right) \geq 0,\quad\quad \mu,\nu\in\{1,2,3\}.
 \label{eq:rea2}
\end{align}
\noindent
Besides constraints on the Reynolds stress tensor itself, it is also possible
to derive constraints on the associated process of evolution, see 
\citep{mishra2014realizability}.

\begin{figure}
\centering
\includegraphics[scale=0.4]{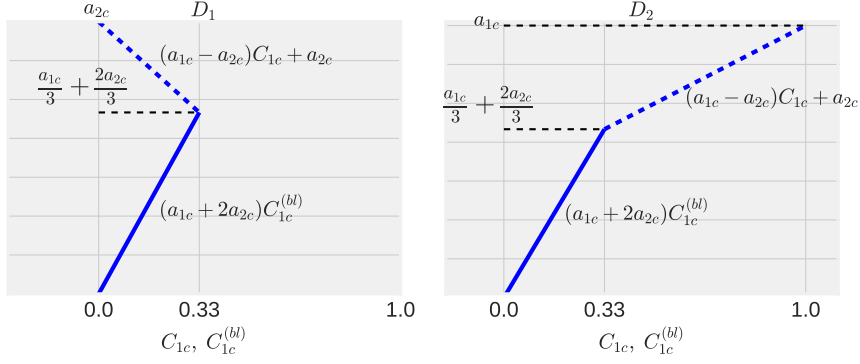}
\caption{The lines of inequality Eq. (\ref{eq:rea_cond_strain_line}) for $D_1$ 
(left, $a_{1c}<a_{2c}$) and $D_2$ (right, $a_{2c}<a_{1c}$). \label{fig:D1_D2}}
\end{figure}

When Eq. (\ref{eq:rea2}) is satisfied, we are guaranteed non-negative diagonal
components and correlations which do not exceed the limits imposed by the
Cauchy-Schwarz inequality. If during the iteration of the turbulence model one
of the diagonal components approaches zero, weak realizability constraints on
its space and time derivative can be enforced in order to avoid 
negative values of this component \citep{pope1985pdf}. To see across
which of the three boundaries of the barycentric map the transition from
positive to negative $\overline{u'_\mu u'_\mu}$ can occur, consider the Reynolds
stress tensor in its principal axes:
\begin{align}
 \widehat{u'_iu'_j} 
 = \frac{2}{3}k\left(
 \begin{tabular}{ccc}
  $2C_{1c} + \frac{1}{2}{C_{2c}} + 1$ & 0 & 0 \\ 
  0 & $-C_{1c} + \frac{1}{2}C_{2c} + 1$ & 0 \\
  0 & 0 & $-C_{1c} - C_{2c} + 1$
 \end{tabular}
 \right),
 \label{eq:rst_pcs}
\end{align}
\noindent
where Eqs. (\ref{eq:b_ij_pcs}), (\ref{eq:b_ij}) and (\ref{eq:bary_coefs}) are 
used to express the diagonal components in terms of the coefficients. We now 
have
\begin{itemize}
\item[] {axi-symmetric contraction:} $C_{1c} =
0\rightarrow \widehat{u'_3u'_3}=0$ at 2C,
\item[] {axi-symmetric expansion:} $\;\;C_{2c} = 0\rightarrow
\widehat{u'_2u'_2} = 0$ and $\widehat{u'_3u'_3}=0$ at 1C,
\item[] {two-component boundary:} $\;C_{3c} =
0\rightarrow$ $\widehat{u'_2u'_2}=0$ at 1C and $\widehat{u'_3u'_3}=0$
everywhere.
\end{itemize}
\noindent
These results show that only along the bottom boundary, is the component 
$\widehat{u'_3u'_3}$ (corresponding to the smallest eigenvalue $\lambda_3$) 
zero, hence the name two-component boundary. Thus, when a principal Reynolds 
stress vanishes somewhere in the flow domain, the corresponding point in the 
barycentric map will be located at the bottom boundary. Depending on the sign of 
the temporal and spatial derivatives of this vanishing $\widehat{u'_\mu 
u'_\mu}$, the trajectory will either cross the boundary yielding unrealizable 
solutions, or move back into the barycentric map. To avoid the former, 
realizability constraints can be imposed on closure models. A convenient 
quantity to impose such constraints is 
\begin{figure}
 \centering
 \includegraphics[scale=0.4]{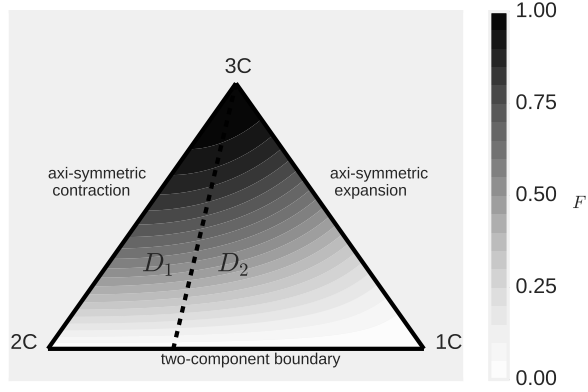}
 \caption{The barycentric map color coded according to
(\ref{eq:Fhat}). Domain $D_1$ and $D_2$ are the two sub triangles
left and right of the plane strain line. \label{fig:F}}
\end{figure}
\begin{align}
 \widehat{F} := \det\left(\frac{\widehat{u'_iu'_j}}{\frac{2}{3}k}\right) =
\left(\frac{2}{3}k\right)^{-3} \cdot \widehat{u'_1u'_1}\cdot \widehat{u'_2 u'_2}
\cdot \widehat{u'_3u'_3},
\label{eq:Fhat}
\end{align}
\noindent
which is defined such that $\widehat{F}=1$ in the case of isotropic turbulence 
and $\widehat{F}=0$ when a principal Reynolds stress vanishes. Thus,
$\widehat{F}\geq0$ holds inside the barycentric map (see Figure \ref{fig:F}),
which can be considered as a normalized version of the third condition in
Eq (\ref{eq:rea2}). The condition of weak realizability ensures
that an initially three-component state of turbulence can never achieve a
two-component state, i.e., where the two-component boundary is made inaccessible
through a constraint on the first derivative of $\widehat{F}$ 
\citep{pope1985pdf}:
\begin{align}
 \left.\frac{\Dif \widehat{F}}{\Dif t}\right|_{\widehat{F}=0} > 0.
 \label{eq:weak_rea}
\end{align}
\noindent
Using Eq. (\ref{eq:rst_pcs}) to expand Eq. (\ref{eq:Fhat}) yields
\begin{align}
 \widehat{F} = 
 (2C_{1c} + \frac{1}{2}{C_{2c}} + 1)(-C_{1c} + \frac{1}{2}C_{2c} + 1)
 (-C_{1c} - C_{2c} + 1)
\end{align}
\noindent
After some algebra (using $C_{2c} = 1 - C_{1c}$ at the two-component boundary), 
we obtain
\begin{align}
\left.\frac{\Dif\widehat{F}}{\Dif t}\right|_{\widehat{F}=0} = 
\left(\frac{9}{4}C_{1c}^2 - \frac{9}{4}\right)\frac{\Dif C_{1c}}{\Dif t}+
\left(\frac{9}{4}C_{1c}^2 - \frac{9}{4}\right)\frac{\Dif C_{2c}}{\Dif t}
\label{eq:rea4}.
\end{align}
\noindent
At this point our transport model Eq. (\ref{eq:model}) for $C_{1c}$ and 
$C_{2c}$ is inserted into Eq. (\ref{eq:rea4}), and the weak realizability 
condition Eq. (\ref{eq:weak_rea}) then becomes
\begin{align}
a_{1c}C_{1c}^{(bl)} + a_{2c}C_{2c}^{(bl)} \leq 
\left(a_{1c} - a_{2c}\right)C_{1c} + a_{2c},\quad\mathrm{when}\;\;\widehat{F}=0.
\label{eq:rea_cond}
\end{align}
\noindent
Under certain conditions described in Section \ref{sec:lag_coefs}, the 
results from a Boussinesq model will lie on top op the plane strain line. Along 
this line $C^{(bl)}_{2c} = 2C^{(bl)}_{1c}$ holds, allowing us to simplify 
Eq. (\ref{eq:rea_cond}) further to 
\begin{align}
 \left(a_{1c} + 2a_{2c}\right)C_{1c}^{(bl)} \leq \left(a_{1c} - 
a_{2c}\right)C_{1c} + a_{2c},\quad\mathrm{when}\;\;\widehat{F}=0.
\label{eq:rea_cond_strain_line}
\end{align}
\noindent
Consider the two domains $D_1, D_2$ at either side of the plane strain line as 
shown in Figure \ref{fig:F}. Let us evaluate the inequality Eq. 
(\ref{eq:rea_cond_strain_line}) in both domains, starting with $D_1$. 
In this domain, $C_{1c}\in\left[0, 1/3\right]$ along the two-component
boundary. Each $C_{1c}$ is associated with a baseline value
$C_{1c}^{(bl)}\in\left[0, 1/3\right]$.
Motivated by the results of Figure \ref{fig:bary_case123}, we assume that 
points are pushed into this domain because we have selected the coefficients 
such that $a_{1c} < a_{2c}$. In this case inequality Eq.
(\ref{eq:rea_cond_strain_line}) is always
satisfied (see Figure \ref{fig:D1_D2}a). Moreover, even if $a_{1c}> a_{2c}$,
Eq. (\ref{eq:rea_cond_strain_line}) will still hold. Note that we assume both
coefficients are positive.

In $D_2$, the range of the baseline coefficient is still
$C_{1c}^{(bl)}\in\left[0,1/3\right]$, but
$C_{1c}\in\left[1/3, 1\right]$ along the two-component boundary. 
If a trajectory enters $D_2$ while $a_{1c} < a_{2c}$, the realizability
condition Eq. (\ref{eq:rea_cond_strain_line}) could be violated.
However, the results of Figure \ref{fig:bary_case123} indicate that for
$a_{1c}<a_{2c}$ the trajectories are pushed away from $D_2$. Hence, under
the condition that $a_{2c} < a_{1c}$, Figure \ref{fig:D1_D2}b shows that on
this side of the plane strain line the model is also weakly realizable. 
%

\begin{figure}
\centering
\includegraphics[scale=0.5]{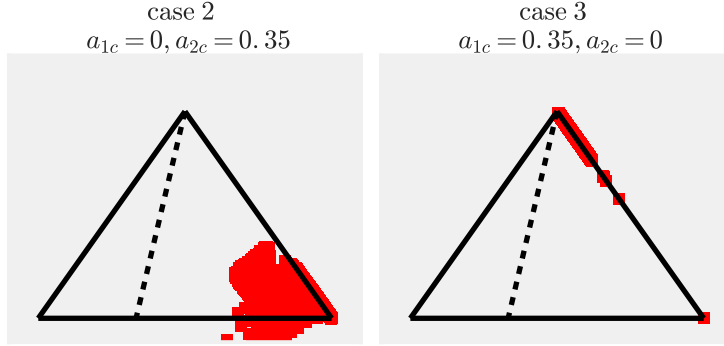}
\caption{Model Eq. (\ref{eq:model}) computed on a
square-duct flow. The boundary conditions were set to 
1C turbulence, i.e., $C_{1c}=1$, $C_{2c}=0$ at the four walls of the duct. 
\label{fig:rea_duct}}
\end{figure}

To test the validity of realizability condition 
Eq. (\ref{eq:rea_cond_strain_line}),
consider a square duct case where the trajectories start at the 
1C corner by setting the wall boundary conditions of model Eq.
(\ref{eq:model}) equal to one-component turbulence. This is the most extreme
case where two principal Reynolds stresses vanish. We then run the model under
the two limiting axi-symmetric cases, i.e., cases 2 and 3 of Section
\ref{sec:lag_coefs}. The preceding analysis suggests that in the latter case the
model should still be realizable, whereas in the former case we can expect
failure. The result are 
shown in Figure \ref{fig:rea_duct}. Note that for case 3 the solution is 
indeed realizable, as the trajectory moves from the 1C to the 3C corner 
along the axi-symmetric expansion border. And as expected, case 2 displays a 
clear violation of realizability. 


\subsection{Perturbation Bounds} \label{sec:bounds}

\begin{figure}
\centering
\includegraphics[scale=0.5]{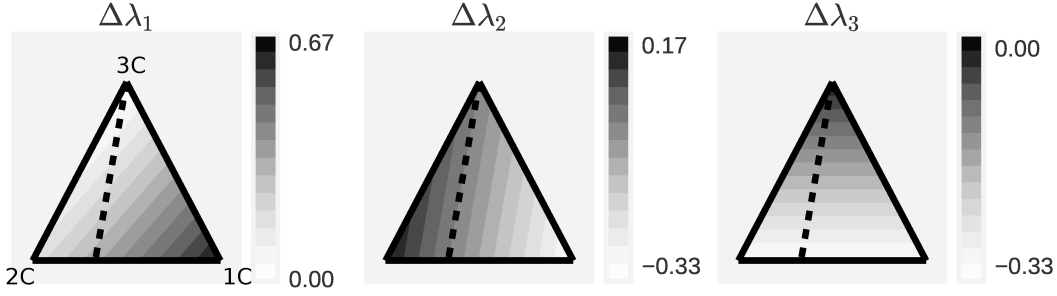}
\caption{Isocontours of eigenvalue perturbations for a 
given value of $C_{1c}^{(bl)}$, which this case is set to zero. 
\label{fig:iso_dLambda}}
\end{figure}

Let us determine the possibility of {\it a priori} identifying the LAG 
coefficients $a_{1c}$ and $a_{2c}$ that will yield the maximum and minimum 
bound on the perturbed eigenvalues that can be achieved with model Eq. 
(\ref{eq:model}). To this end, we decompose the anisotropy tensor as
\begin{align}
 b_{ij} = b_{ij}^{(bl)} + v_{ik}\check{E}_{kl}v_{jl}.
 \label{eq:b_decomp}
\end{align}
Here, $\check{E}_{ij}$ represents the perturbation tensor and is given by 
$\check{E}_{ij}:=\mathrm{diag\left(\Delta\lambda_1, \Delta\lambda_2, 
\Delta\lambda_3\right)}$,
%
where $\Delta\lambda_\alpha$ is the perturbation such that $\lambda_\alpha = 
\lambda^{(bl)}_\alpha + \Delta\lambda_\alpha$. Decomposition Eq.
(\ref{eq:b_decomp}) is possible because $b_{ij}$, 
$b_{ij}^{(bl)}$ and $\check{E}_{ij}$ share the same eigenspace. Note that the 
superscript of $\check{E}_{ij}$ indicates that the increasing diagonal order 
found in $\hat{b}_{ij}$ might not be present in $\check{E}_{ij}$. Assuming 
again that the baseline model will follow the plane strain line in the 
barycentric map, $\check{E}_{ij}$ can be written as
\begin{align}
 \check{E}_{ij} = \left(
 \begin{tabular}{ccc}
  $\frac{2}{3}\left[C_{1c} - \frac{3}{2}C_{1c}^{(bl)}\right] + 
   \frac{1}{6}C_{2c}$ & 0 & 0 \\
  0 & $-\frac{1}{3}C_{1c} + \frac{1}{6}C_{2c}$ & 0 \\
  0 & 0 & $-\frac{1}{3}\left[C_{1c} - 3C_{1c}^{(bl)}\right] - \frac{1}{3}C_{2c}$
 \end{tabular}
 \right).
 \label{eq:E_ij2}
\end{align}
\noindent
By fixing the baseline coefficient to an arbitrary value 
$C_{1c}^{(bl)}\in\left[0,1/3\right]$, we can plot the isocontours of the 
$\Delta\lambda_\alpha$ in order to identify the directions of maximum and 
minimum perturbation. The results are shown in Figure \ref{fig:iso_dLambda}. 
Clearly, $\Delta\lambda_1$ is maximized in the direction of 1C, and minimized 
towards 3C. The perturbation $\Delta\lambda_2$ is maximized at 2C and minimized 
at 1C. Finally, the maximum of $\Delta\lambda_3$ is located at 3C, and its 
minimum is found anywhere along the 2C boundary. Thus, straight lines in the 
direction of minimum/maximum $\Delta\lambda_\alpha$ are the axi-symmetric 
expansion, contraction and the plane-strain line. Trajectories along these 
lines can be obtained with {\it a priori} known values for $a_{1c}$ and 
$a_{2c}$ (see Figure \ref{fig:bary_case123}).


\section{Results} \label{sec:res}

\subsection{QoI Bounds} \label{sec:res_plate}

\begin{figure}
\centering
\includegraphics[scale=0.5]{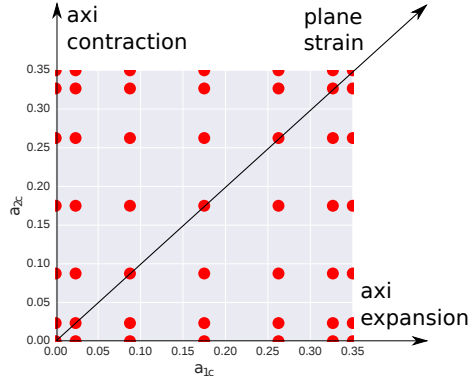}
\caption{A tensor product of 1D Clenshaw-Curtis abscissas
$a_{\alpha c} = -\cos\frac{(j-1)\pi}{m-1}$, $j=1,\cdots, m$. The considered 1D 
range is $\left[0, 0.35\right]$.
\label{fig:CC_grid}}
\end{figure}

Let us consider a flat plate flow with a Reynolds number of $\mathrm{Re} = 
1.1\times10^6$ based on the plate length. To examine the bounds on the QoIs 
resulting from using model Eq.
(\ref{eq:model}), we construct a two-dimensional tensor product
of LAG coefficient from one-dimensional Clenshaw-Curtis abscissas (see Figure 
\ref{fig:CC_grid}). Next, we simply
run the model on each $(a_{1c}, a_{2c})$ combination and examine the possible
range of solutions. Note that one could use the samples obtained in this way
towards constructing a stochastic collocation model 
\citep{witteveen2009uncertainty}.
\begin{figure}
 \centering
 \includegraphics[scale=0.7]{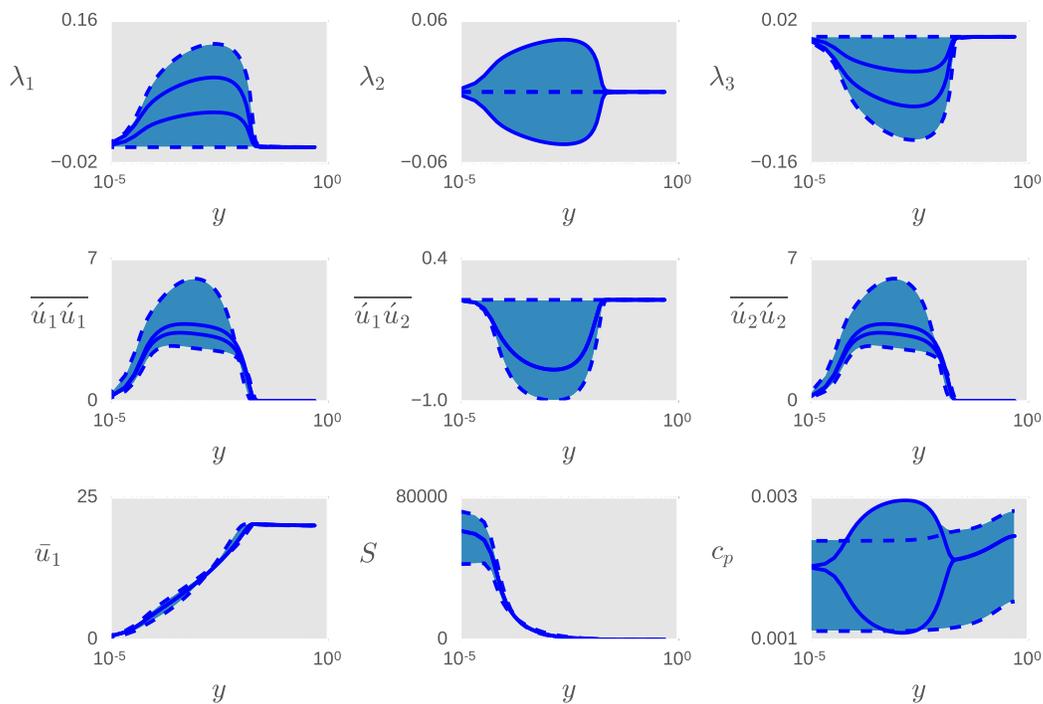}
 \caption{The three anisotropy eigenvalues, Reynolds-stress components, mean 
streamwise velocity, mean strain-rate magnitude ($S:=\sqrt{S_{ij}S_{ji}}$) and
the pressure coefficient versus the wall-normal direction in log scale, at 
streamwise location $x = 0.8$. The dashed lines
are the solutions at the opposite ends along the
plane-strain line (computed using $(0,0)$ and $(0.35, 0.35)$), and the
solid lines denote the axi-symmetric profiles ($(0,
0.35)$ and $(0.35, 0)$).\label{fig:bounds_plate}}
\end{figure}

The results are shown in Figure \ref{fig:bounds_plate}. Notice that most QoI 
are bounded by a plane-strain bound, denoted by the dashed lines. The first
exception is $\lambda_2$, which is zero along the plane-strain line by
definition. Instead,  $\lambda_2$ is bounded by the
axi-symmetric profiles. Moreover, these profiles are symmetric around
$\lambda_2 = 0$. The other exception is the pressure coefficient
$c_p:={(p - p_\infty)}/{(\rho V_\infty^2/2)}$ (in wall-normal direction), which
is bounded by both the plane-strain and axi-symmetric results.

Let us reiterate here that the bounds of Figure
\ref{fig:bounds_plate} are a representation of the uncertainty in the shape of
$\rst$ only. To inject additional uncertainty into the problem, perturbations in
the orientation of the Reynolds stress tensor should be included (see, e.g.,
\citep{mishra2016enveloping}).


\section{Conclusions} 

We introduced two model transport equations for the coefficients of the 
barycentric map to find intervals due to epistemic 
model-form uncertainty in RANS closures. The output of this model is used to 
compute spatially varying perturbations of the eigenvalues 
of the anisotropy tensor. By incorporating so-called LAG terms, these 
perturbations are made with respect to a baseline eddy-viscosity model, such
that in certain areas the results of this underlying model are
recovered. The global behavior of the discrepancy model is governed by two LAG
coefficients, the difference in which determines the amount of perturbation from
the baseline state. In which direction of the barycentric map the perturbations
are made is determined by the sign of the difference. Moreover, weak
realizability conditions and bounds for the eigenvalue perturbations can also be
associated with certain {\it a priori} known values of the LAG coefficients.
This association results in an intuitive discrepancy model, amenable to physical
reasoning. 

When applying the model to a turbulent flat plate flow, we found that the 
Quantities of Interest (QoIs) are bounded by the same {\it a priori} known 
coefficient values. Note that we make no claim as to the generality of this 
result. 
Future work will involve computing other flow cases, as well as a Bayesian 
analusys in order to find stochastic (posterior) estimates of the LAG 
coefficients and 
QoIs. Thus, the results of this paper can be considered as a first step
towards a full Bayesian description of the uncertainty in the eigenvalues, in 
the sense that it describes the effects of (uniform) prior distributions on the
model coefficients. A further reseach option for the future is the
investigation of non-linear return to eddy viscosity models. 

\section*{Acknowledgments} 

This investigation was funded by the United States Department of Energy’s (DoE) 
National Nuclear Security Administration (NNSA) under the Predictive Science 
Academic Alliance Program II (PSAAP II) at Stanford University.

\bibliographystyle{ctr}

\end{document}